\documentclass[prd,,aps,nofootinbib,amsmath,amssymb]{revtex4}
\usepackage{amssymb,amsmath,epsfig,bm}
\usepackage[usenames,dvipsnames]{color}
\usepackage[bookmarks]{hyperref}
\usepackage{mathrsfs}

\newcommand{\Scri}{\mathscr{I}}

\begin{document}

\title{Observational distinction between black holes and naked singularities: the role of the redshift function}
\author{N\'estor Ortiz$^{1,2}$, Olivier Sarbach$^{1,2}$, and Thomas Zannias$^{1,3}$}

\affiliation{
$^1$Instituto de F\'{\i}sica y Matem\'aticas,
Universidad Michoacana de San Nicol\'as de Hidalgo,\\
Edificio C-3, Ciudad Universitaria, 58040 Morelia, Michoac\'an, M\'exico.\\
$^2$Perimeter Institute for Theoretical Physics, 31 Caroline Street, Waterloo, Ontario N2L 2Y5, Canada.\\
$^3$Department of Physics, QueenÕs University, Kingston, Ontario K7L 3N6, Canada.}
\email{nortiz@perimeterinstitute.ca, sarbach@ifm.umich.mx, zannias@ifm.umich.mx}

\begin{abstract}
We suggest that the redshift of photons traveling from past to future null infinity through a collapsing object could provide an observational signature capable of differentiating between the formation of a globally naked singularity and the formation of an event horizon. Supporting evidence for this idea is drawn from the analysis of photons with zero angular momentum through the center of a collapsing spherical dust cloud. We show that the frequency shift as a function of proper time with respect to stationary observers has distinct features depending on whether the object collapses to a black hole or a naked singularity.
\end{abstract}

\maketitle

{\it Introduction.}
The possibility that nature admits the formation of stable naked singularities as a result of the complete gravitational collapse of a bounded system has not been ruled out so far. Indeed, a general proof or disproof of the weak cosmic censorship conjecture~\cite{rP69} still constitutes a challenging problem for theoretical physics~\cite{rW97,rP99}. However, it has been shown that naked singularities appear within simplified collapse models with a high degree of symmetry. For instance, this is the case for the collapse of a spherical dust cloud~\cite{dElS79,dC84,rN86,pJiD93,Joshi-Book} or the spherical collapse of a massless scalar field~\cite{dC94,mC93}. Although the naked singularities in the latter case are harmless in the sense that they have been shown to be unstable~\cite{dC99}, the situation in the dust collapse case is not so clear. Unlike the scalar field case, the naked singularities encountered in the dust model are stable under spherical perturbations~\cite{dC84,fMrTpJ00,nOoS11}. A highly relevant question is whether or not they are also stable under nonspherical perturbations. For progress on this important issue, see for instance Refs.~\cite{bWkL89,hItHkN98,hItHkN99,hItHkN00,eDbN11b,eDbN11,nOoS14,nOoS16}. 

From the theoretical and in particular the astrophysical point of view, it is important to have {\it ready made theoretical tools} capable of differentiating the formation of a naked singularity from the formation of an event horizon during the collapse of a bounded system. Important information regarding the nature of a collapsing object is revealed from the study of the spectrum of electromagnetic radiation emitted from the surface or the interior of the object, see for example Refs.~\cite{wAkT68,jJ69,kLrR79,hLvF06,vFkKhL07,lKdMcB14,lKdMcB15}. 
In this work, following~\cite{nOoS14b}, we consider a different physical scenario in which the radiation, instead of being generated from the interior of the collapsing object is originated from a distant source and crosses the collapsing object without interacting with the matter. For instance, this radiation could be emitted from a distant supernova explosion or be part of the cosmic microwave background. Our scenario has some similarities with the idea based on gravitational lensing in~\cite{kVdNsC98,kVgE02,kVcK08} aiming to distinguish the naked singularity described by a static, spherically symmetric solution of the Einstein massless scalar field equations from a black hole, and also with the recent proposal in~\cite{pJdMrN14}. However, in contrast to the approaches in~\cite{kVdNsC98,kVgE02,kVcK08,pJdMrN14}, the present work exploits the dynamical phase of the collapse and is focused on the behavior of incoming photons grazing the singularity. Furthermore, we claim that the naked singularities considered in our work are physically more realistic than the ones considered in~\cite{kVdNsC98,kVgE02,kVcK08}. In our model, the singular end state is formed from regular initial data; see also~\cite{pJdMrN14} for an alternative idea to generate a singular end state from regular data.

In this Note, we analyze our scenario for the particular case of a collapsing, spherical dust cloud and photons with vanishing angular momentum. Specifically, we consider photons traveling from past to future null infinity through the center of the collapsing dust cloud and show that the redshift experienced by these photons as a function of proper time of a distant stationary observer exhibits distinct features depending on the final fate of the cloud. The more generic case of photons with nonvanishing angular momenta is analyzed in a companion paper~\cite{nOoStZ15}.

\begin{figure}[]
\begin{center}
\includegraphics[width=8.5cm]{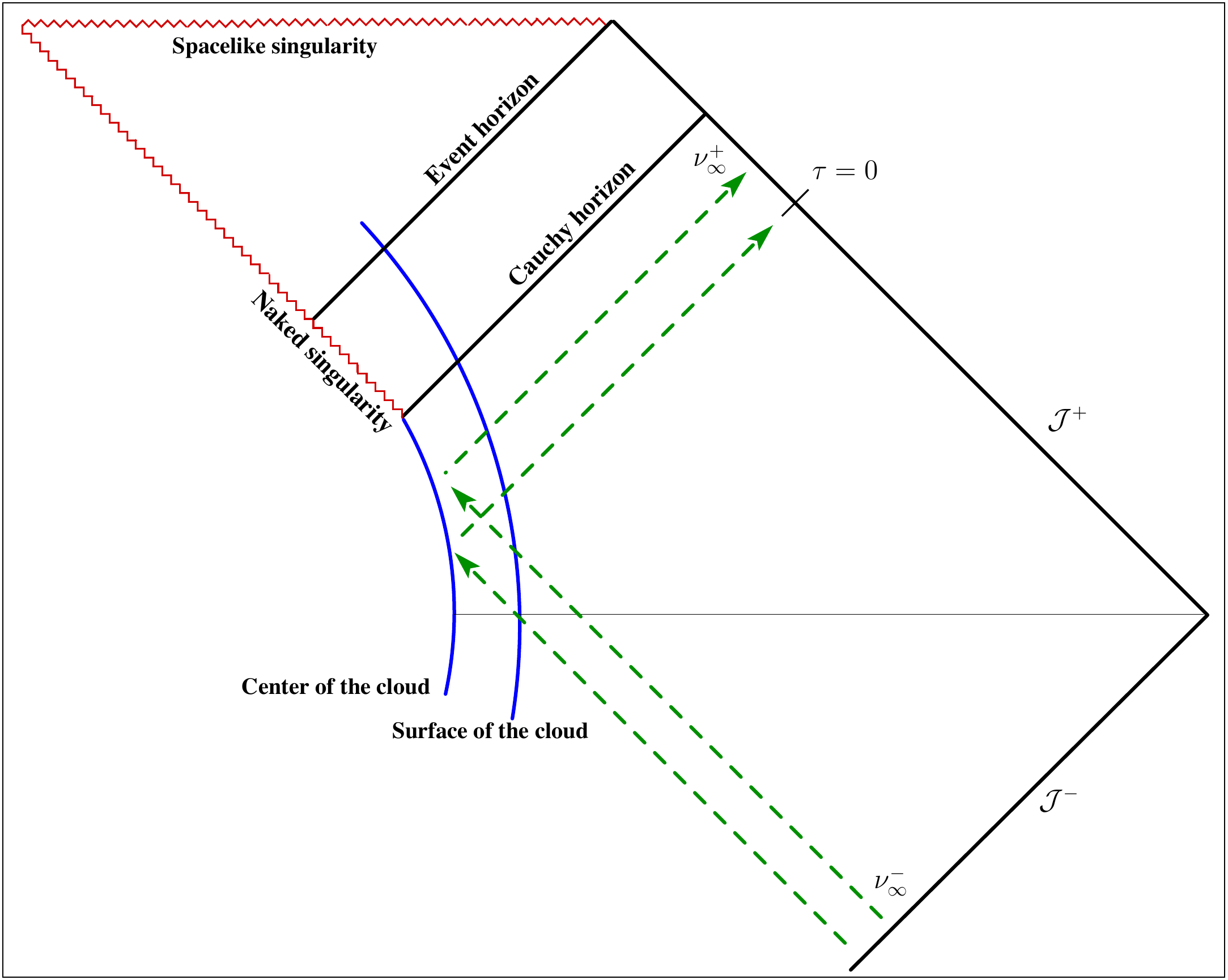}
\end{center}
\caption{\label{Fig:Conformal} Conformal diagram illustrating the physical situation under consideration: a beam of photons propagating from $\Scri^-$ to $\Scri^+$ through the center of the cloud. The frequency of these photons when measured by asymptotic static observers in the past (future) is $\nu_\infty^-$ ($\nu_\infty^+$). The light ray labeled by $\tau=0$ corresponds to the one penetrating the cloud at the moment of time symmetry (when the cloud is momentarily at rest, indicated by the horizontal line), and marks the moment at which observers in the future asymptotic region turn on their detectors.}
\end{figure}

{\it Observational features of naked singularities from the total redshift.}
We consider the interior of a collapsing, spherical dust cloud of finite initial radius, described by the Tolman-Bondi metric, see for instance~\cite{MTW-Book}. The exterior of the cloud is described by the Schwarzschild metric. A beam of photons with zero angular momentum is injected from past null infinity ($\Scri^-$) having frequency $\nu_\infty^-$ as measured by asymptotic static observers. This beam traverses the collapsing cloud through its center and finally reaches future null infinity ($\Scri^+$) with frequency $\nu_\infty^+$ as measured by asymptotic static observers; see Fig.~\ref{Fig:Conformal}.

We assume that the dust cloud is at rest at some initial time $t=0$, and model the initial density profile by the following family of functions~\cite{nOoS11}
\begin{equation}
\rho(R) = \rho_c\times\left\{ \begin{array}{ll} 
 1 - 2a\left(\frac{R}{R_1}\right)^2 + (2a-1)\left(\frac{R}{R_1}\right)^4, &
0 \leq R \leq R_1,\\
0,&
R > R_1, \end{array} \right.
\label{Eq:rho0}
\end{equation}
where $R_1 > 0$ is the initial areal radius of the cloud, $\rho_c > 0$ is the central density, and $a$ is a dimensionless parameter restricted to the interval $0 < a \leq 1$ which characterizes the flatness of the profile $\rho$. The initial density defined in Eq.~(\ref{Eq:rho0}) is continuous and monotonously decreasing on the interval $[0,\infty)$, and except for the lack of differentiability at $R = R_1$, it satisfies all the assumptions (i)--(viii) listed in Sec.~II of Ref.~\cite{nOoS11} such as smoothness, boundedness, nonnegative mass density, absence of shell-crossing singularities, absence of initially trapped surfaces, et cetera. In particular, the profile given in Eq.~(\ref{Eq:rho0}) is smooth at the center, such that $\partial_R \rho(0) = 0$, and furthermore it satisfies $\partial^2_R\rho(0) < 0$. As shown by Christodoulou~\cite{dC84}, these conditions lead to the formation of a shell-focusing singularity which is visible at least to local observers. In Ref.~\cite{nOoS11} we discussed the conditions upon the parameters $R_1$, $\rho_c$ and $a$ leading to the formation of a globally naked singularity.

The initial compactness ratio ${\cal C} := 2m_1/R_1$ of a collapsing cloud with initial density given by Eq.~(\ref{Eq:rho0}) turns out to be
\begin{equation}\label{Eq:Total_mass}
{\cal C} = \frac{8}{3}\pi R_1^2\rho_c \left[ 1 - \frac{6}{5}a + \frac{3}{7}(2a - 1) \right].
\end{equation}
For simplicity, in our numerical simulations we fix $a = 0.5$, nonetheless other values in the interval $(0,1]$ lead to the same qualitative results. Next, based on the causal structure of the collapsing spacetime revealed by the numerical algorithm introduced in Ref.~\cite{nOoS11} for the construction of conformal diagrams, we make the particular choice ${\cal C} = 0.2 =: {\cal C}_0$. This choice leads to the formation of a black hole, whereas compactness ratios ${\cal C}$ smaller than or equal to ${\cal C}_0/2$ lead to the formation of a globally naked singularity.

We numerically compute the frequency shift $\nu_\infty^+ / \nu_\infty^-$ as a function of proper time $\tau$ of a distant static observer. The observer starts his chronometer ($\tau = 0$) when he receives the particular radial light ray originating from $\Scri^-$ and penetrating the cloud at time $t=0$, when the cloud is at rest, see Fig.~\ref{Fig:Conformal}. The explicit formula for computing this frequency shift function has been derived in Ref.~\cite{nOoS14b}. The numerical integration along the radial light rays required for this computation was performed using a fourth-order Runge-Kutta algorithm. Our numerical solutions were checked to be self-convergent to fourth order accuracy and consistent with the analytic results for the case of homogeneous density distributions~\cite{nOoS14b}.

\begin{figure}[h!]
\begin{center}
\includegraphics[width=9.1cm]{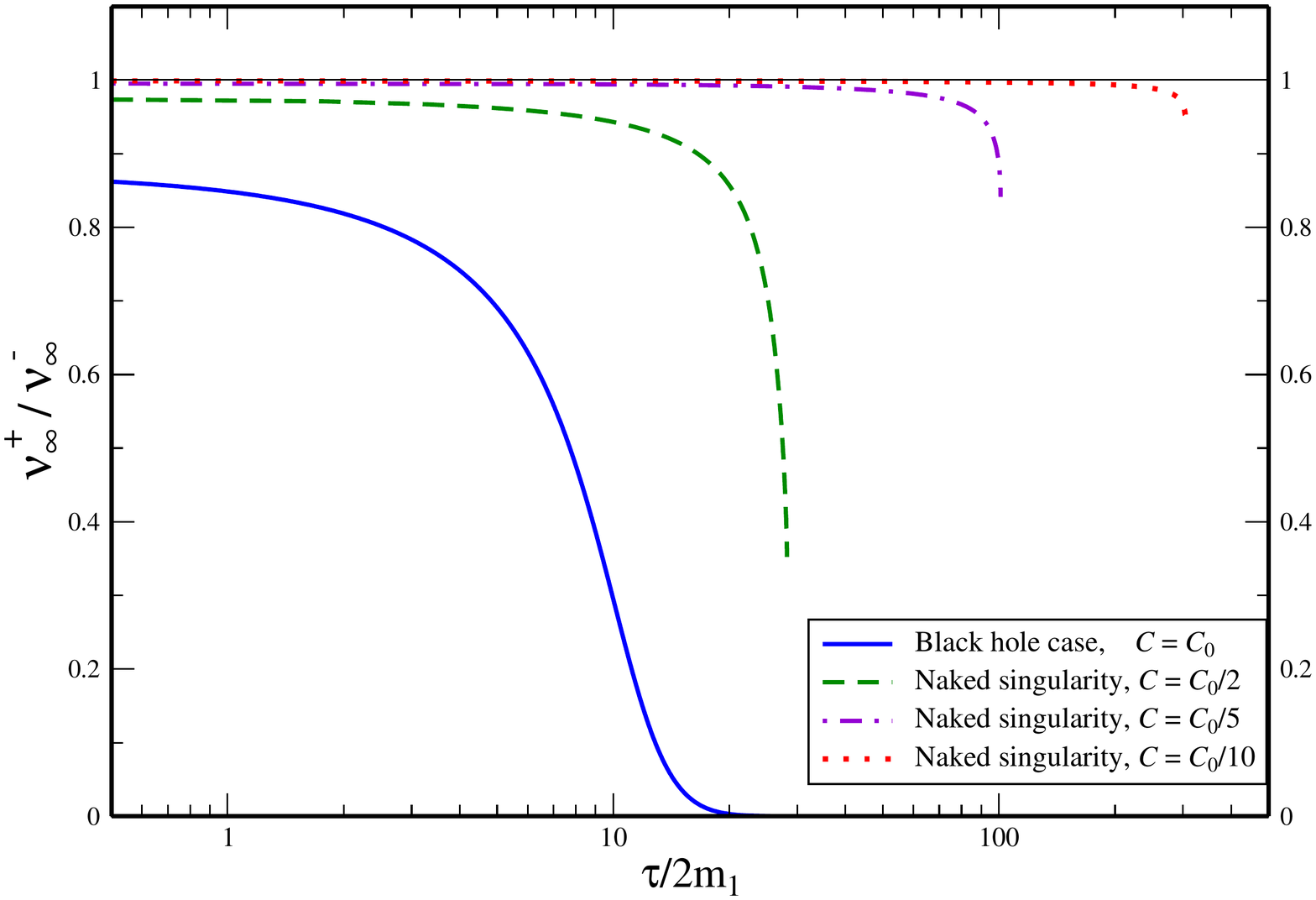}
\hspace{-0.7cm}
\includegraphics[width=9.1cm]{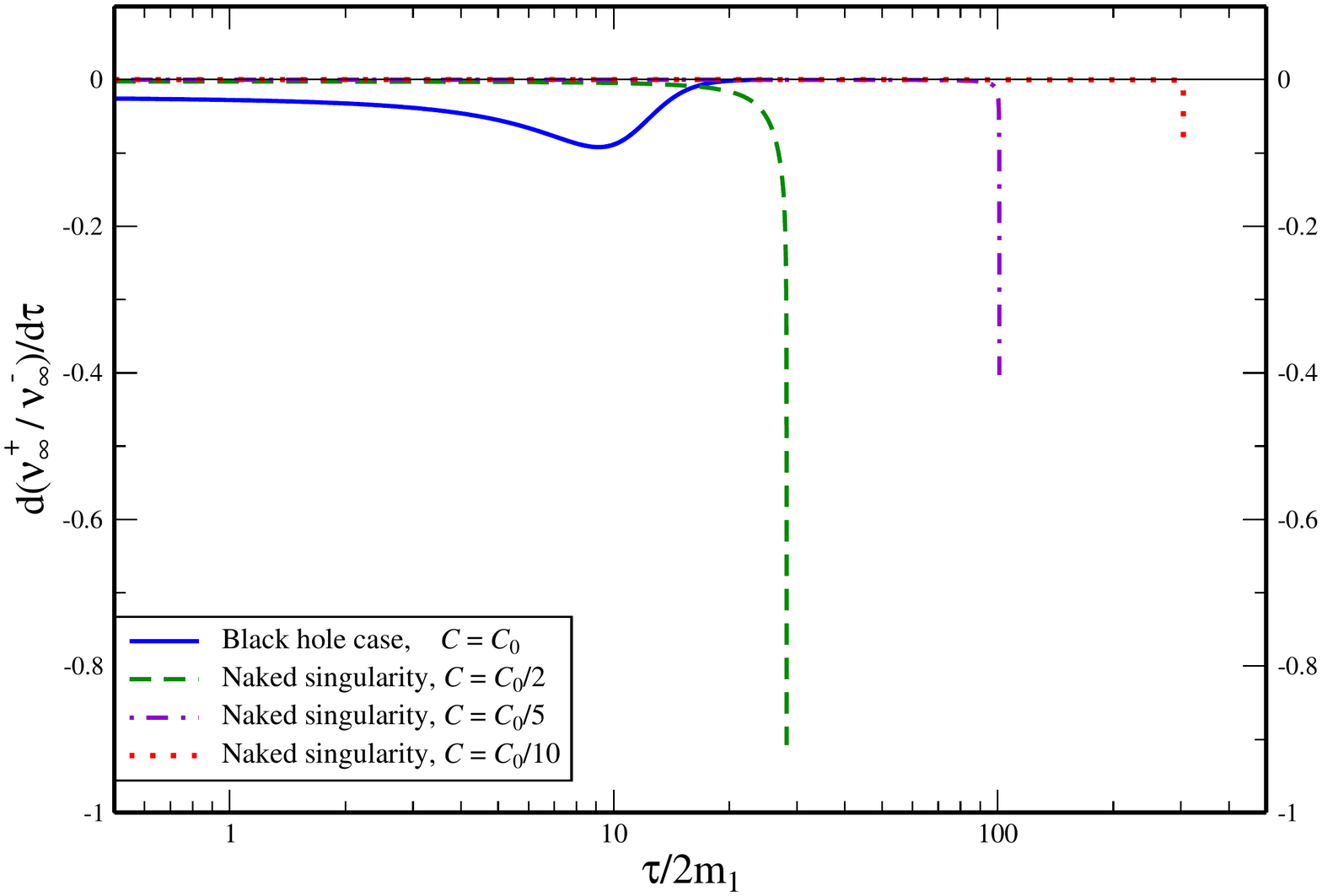}
\end{center}
\caption{\label{Fig:naked} Left panel: comparison of frequency shifts in the cases of black hole formation and naked singularities resulting from the gravitational collapse of a spherical dust cloud with the density profile described in Eq.~(\ref{Eq:rho0}) with the flatness parameter $a=0.5$. The solid line corresponds to a dust cloud of initial compactness ratio ${\cal C} = {\cal C}_0 = 0.2$ forming a black hole. The dashed lines correspond to collapsing clouds with smaller values of ${\cal C}$ giving rise to naked singularities. Right panel: time derivative of the frequency shifts shown in the left panel. We observe that the first and second time derivatives of these curves exhibit a qualitatively different behavior in the black hole case than in the case where a naked singularity forms.
}
\end{figure}

The left panel of Fig.~\ref{Fig:naked} shows the total frequency shift $\nu_\infty^+/\nu_\infty^-$ as a function of $\tau$. The solid line corresponds to a cloud with initial compactness ratio ${\cal C} = {\cal C}_0$ collapsing to a black hole. The dashed lines correspond to clouds with smaller values of ${\cal C}$ giving rise to naked singularities. The numerical integration of light rays stops just a moment before the formation of either an event horizon associated with a black hole, or a Cauchy horizon associated with a naked singularity. We observe that in the black hole case, the photons experience an infinite redshift since $\nu_\infty^+/\nu_\infty^- \to 0$ as they approach the event horizon, as expected from the properties of the Schwarzschild spacetime. In sharp contrast, the sudden cut-off of the frequency shift as the photons approach a Cauchy horizon shows that the redshift is finite, confirming the theoretical prediction given in Ref.~\cite{nOoS14}.\footnote{For the corresponding result in the case of marginally bound collapse, see Ref.~\cite{iD98}.} Note that this cut-off is due to the fact that after some time, light rays originating from $\Scri^-$ do not any longer pass through the center of the cloud and reach $\Scri^+$; instead they terminate at the spacelike portion of the singularity, see Fig.~\ref{Fig:Conformal}. This cut-off represents the first important observable feature that can help recognize a naked singularity.

The right panel of Fig.~\ref{Fig:naked} shows the first derivative of the frequency shift $\nu_\infty^+/\nu_\infty^-$ with respect to proper time $\tau$ of a distant static observer. We note that this time derivative reaches a finite negative value as the observer approaches the Cauchy horizon; however this behavior is essentially different in the black hole case, where the time derivative of the frequency shift smoothly vanishes as the observer approaches the event horizon. This feature constitutes the second qualitative distinction between naked singularities and black holes. Finally, from the right panel of Fig.~\ref{Fig:naked} we infer that the second derivative of the frequency shift also plays an important role in the distinction between naked singularities and black holes: in the former case, the second derivative remains negative and tends to diverge as the observer approaches the Cauchy horizon, while in the black hole case it eventually becomes positive and converges to zero as the observer approaches the event horizon.
These three observable features suggest a systematic procedure to discriminate a naked singularity from a black hole based on observations of the redshift function.

\begin{figure}[h!]
\begin{center}
\includegraphics[width=9.1cm]{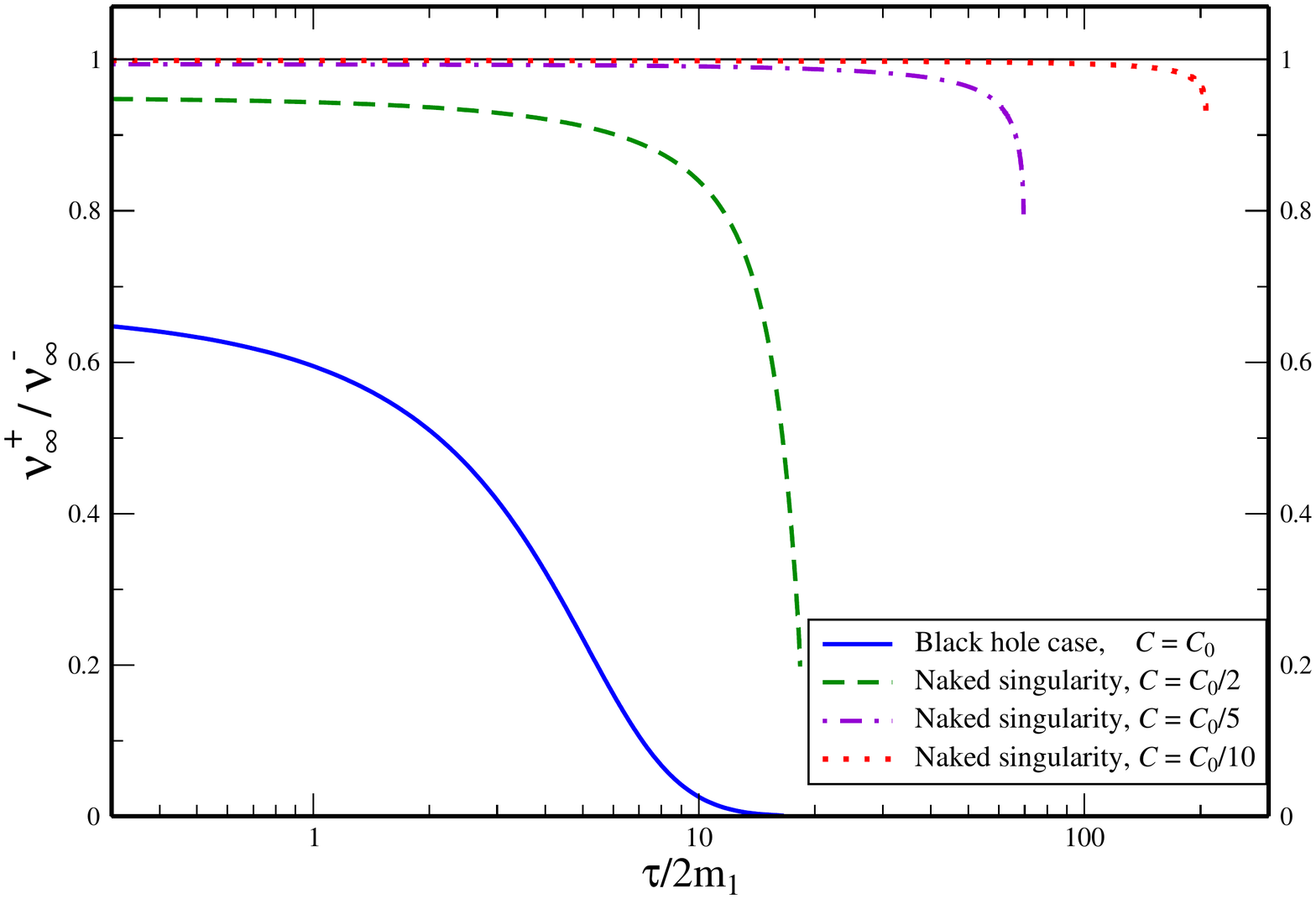}
\hspace{-0.7cm}
\includegraphics[width=9.1cm]{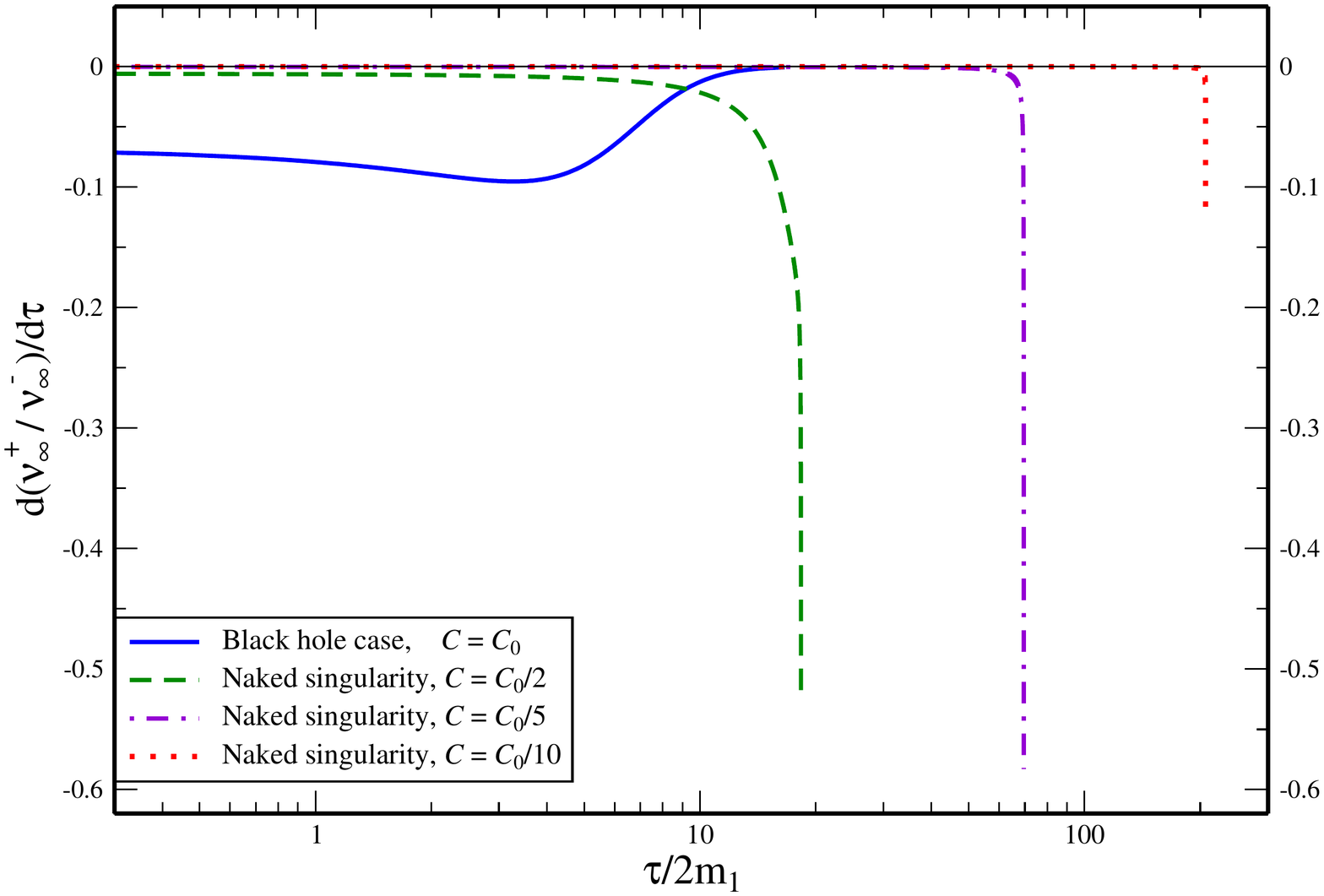}
\end{center}
\caption{\label{Fig:naked_smooth} The same functions as in Fig.~\ref{Fig:naked}, employing the smoother density profile defined in Eq.~(\ref{Eq:rho_smooth}) with $a_0 = 1$ and ${\cal C}_0 = 16/77$. We observe that the features shown in Fig.~\ref{Fig:naked} are preserved in this scenario, suggesting that our results are robust.}
\end{figure}

It is important to show that the qualitative properties discussed so far do not depend on the particular features of the density profile employed in Eq.~(\ref{Eq:rho0}). In order to provide supportive evidence for this statement, we repeat the numerical computations discussed above using the smoother family of density profiles given by
\begin{equation}\label{Eq:rho_smooth}
\rho(R) = \rho_c\times\left\{ \begin{array}{ll} 
 1 - \frac{5}{3}a_0\left( \frac{R}{R_1} \right)^2 + \frac{7}{3}a_1 \left( \frac{R}{R_1} \right)^4 + 3a_2 \left( \frac{R}{R_1} \right)^6 + \frac{11}{3}a_3\left( \frac{R}{R_1} \right)^8, & 0 \leq R \leq R_1,\\
0,& R > R_1. \end{array} \right.
\end{equation}
As before, $R_1 > 0$ is the initial areal radius of the cloud, $\rho_c > 0$ the central density, and the dimensionless parameter $a_0$ characterizes the flatness of the profile. The remaining parameters are determined by $a_1 = -3(6 - 5a_0)/7$, $a_2 = (8-5a_0)/3$ and $a_3 = -(9 - 5a_0)/11$. Presently, $a_0$ is required to fulfill the inequality $0 <  a_0 < 12/5$ in order to satisfy the conditions (i)--(viii) listed in Sec.~II of Ref.~\cite{nOoS11} with the exception of $C^\infty$-differentiability at $R = R_1$. As in the previous case, this density profile is continuous and monotonously decreasing on the interval $[0,\infty)$. However, in contrast to the previous case, this profile is twice continuously differentiable everywhere. The initial compactness ratio of the cloud described by the profile defined in Eq.~(\ref{Eq:rho_smooth}) is given by
\begin{equation}\label{Eq:mass_smooth}
{\cal C} = \frac{8}{3}\pi R_1^2\rho_c \left( 1 - a_0 + a_1+ a_2 + a_3 \right).
\end{equation}
Again, based on the numerical algorithm described in Ref.~\cite{nOoS11}, we know that ${\cal C} = 16/77 =: {\cal C}_0$ leads to the formation of a black hole, whereas ${\cal C} \leq {\cal C}_0/2$ guarantees the appearance of a globally naked singularity.
In Fig.~\ref{Fig:naked_smooth} we plot the same quantities as in Fig.~\ref{Fig:naked} employing the density profile defined in Eq.~(\ref{Eq:rho_smooth}) with the choice $a_0 = 1$. Clearly, the observable qualitative differences between naked singularities and black holes persist.

{\it Discussion.}
We have studied the frequency shift of light rays traveling on an asymptotically flat dynamical spacetime describing the complete gravitational collapse of a spherically symmetric dust cloud. Specifically, we considered photons propagating from past to future null infinity through the center of the cloud and discussed how the information obtained from the frequency shift measured by distant stationary observers can help discriminate naked singularities from black holes. Our analysis revealed the following three fundamental features exhibited by the  frequency shift as a function of proper time, see Figs.~\ref{Fig:Conformal},\ref{Fig:naked},~and~\ref{Fig:naked_smooth}:
$(i)$ the frequency shift corresponding to outgoing photons propagating in the vicinity of the Cauchy horizon associated with a naked singularity shows a sudden cut-off with a finite redshift. In contrasts to this, in the black hole case, the frequency shift goes smoothly to zero, and thus the corresponding redshift factor diverges;
$(ii)$ the first time derivative of the photons' frequency shift behaves in a completely different way in the vicinity of a Cauchy horizon associated with a naked singularity than in the black hole case. In the first case this derivative reaches a finite negative value as the observer approaches the Cauchy horizon, while in the second case the derivative vanishes as the observer approaches the event horizon;
$(iii)$ the analysis of the behavior of the second time derivative of the frequency shift shows a clear difference between the naked singularity and the black hole case. For the case of the naked singularity, this derivative remains always negative and tends to diverge as the observer reaches the Cauchy horizon, in contrast to the black hole case, where it exhibits a change of sign and vanishes as the observer reaches the event horizon, see the right panels of Figs.~\ref{Fig:naked}~and~\ref{Fig:naked_smooth}. We have verified that the qualitative distinctions $(i)-(iii)$ are robust in the sense that they persist when different initial data are considered. Therefore, we conclude that within our model it is, in principle, possible to differentiate the formation of a naked singularity from the formation of a black hole based on the detection of photons traversing a collapsing object.

However, it should be stressed that a more definite confirmation regarding the possibility of differentiating the formation of black holes from the formation of naked singularities requires additional work. For instance, it is important to generalize this work to light rays which do not necessarily traverse the cloud through its center. In~\cite{nOoStZ15} we extend our analysis to the case of photons with nonvanishing angular momenta. Apart from confirming the validity of our results, the inclusion of angular momenta has several interesting consequences on the image of the source perceived by the asymptotic observer, see~\cite{nOoStZ15}. In addition to the generalization to non-radial photons, the collapse model should also be made more realistic, and effects due to pressure, rotation and magnetic fields should be considered. 

We should also mention that the geometric optics approximation employed in this work implies that photons are moving on future directed null geodesics (or light rays in the terminology of this approximation) and that they propagate freely through the collapsing medium. Although for the case of the event horizon formation this approximation is expected to be reliable, since all interactions take place away from extreme conditions of unbounded curvature and matter density, that is not any longer the case when a Cauchy horizon forms. In the presence of high curvature and matter density, one expects photon trapping to take place, or for instance, Compton scattering to become important. Although the effects of these processes ought to be analyzed in detail, it is expected that the abrupt cut off in the frequency shift as detected by the asymptotic observer is preserved since it is a feature of the geometry of the collapse and the latter is independent of the interaction mode between the collapsing matter and the illuminating radiation. It is our hope that future investigations will clarify some of these open issues.

We conclude this note with some comments regarding the feasibility of applying the predictions of this work to decide whether a black hole or a naked singularity is the outcome of a complete gravitational collapse. It requires a collapsing configuration, asymptotic observers, and illuminating sources. In a stellar collapse within our galaxy, the illuminating external radiation could be generated by the other stars. For cosmological settings, a type Ia-cosmological supernovae could illuminate a collapsing spherical dust inhomogeneity leading to the formation of a supermassive black hole (or to a naked singularity).\footnote{However, in such scenarios, one should also take into account the effects from the ambient cosmological expansion.} We remark however, that in our collapse model the time scale for which the redshift variations should be detected is of the order of
\begin{equation}
10 \times 2m_1 \approx 10^{-4}\left( \frac{M}{M_\odot} \right) sec
\end{equation}
in the black hole case, while depending on the initial compactness ratio of the cloud it could be higher by one or more orders of magnitude in the naked singularity case, see Figs.~\ref{Fig:naked}~and~\ref{Fig:naked_smooth}. For stellar mass black holes, whose masses $M$ range between $1-100$ solar masses $M_\odot$, this time scale is probably too short to be resolved with current techniques. However, much larger time scales are obtained 
for collapse leading to supermassive black holes.

Finally, it is important to stress that  the observability of the redshift variations predicted by our work, would necessitate instruments capable of resolving sources exhibiting angular scales at the level of microarcseconds, as is the case, for instance, for Sagittarius A$^*$, the supermassive black hole in the center of our galaxy~\cite{sDetal08}, which has $M\simeq 4\times 10^6 M_\odot$. Interestingly, nowadays millimeter-wave very-long baseline interferometric arrays such as the Event Horizon Telescope~\cite{EHT} are being able to resolve the region around Sagittarius A$^*$ to scales of the order of its gravitational radius. These technological advances give hope that our redshift effect could be measured in the near future.

\acknowledgments
This work was supported in part by CONACyT Grants No. 232390, 46521, 101353, and 234571  and by a CIC Grant to Universidad Michoacana. Research at Perimeter Institute is supported by the Government of Canada through Industry Canada and by the Province of Ontario through the Ministry of Research and Innovation.

\bibliographystyle{unsrt}
\bibliography{../../References/refs_collapse}

\end{document}